\documentclass{aa}  

\usepackage{graphicx}
\usepackage{txfonts}
\usepackage{lipsum}
\usepackage{subcaption}         % necessary for continued figures, example in section 3
\usepackage{lscape}             % to rotate a single page table, example in appendix.
\usepackage{placeins}           % useful with \FloatBarrier, to keep 
\usepackage{hyperref}
\hypersetup{
    colorlinks=true,
    linkcolor=blue,
    citecolor=blue,
    urlcolor=blue
}                              
\begin{document}

%%%%%%%%%%%%%%%%%%%%%%%%%%%%%%%%%%%%%%%%
% if you use custom commands in your title,
% ensure to check your title when submitting!
%%%%%%%%%%%%%%%%%%%%%%%%%%%%%%%%%%%%%%%%
   % \title{\texttt{catcosmo}: A Study of Cat-Conditioned Random Seeds and Their Impact on Deep Learning-Based Galaxy Classification}
\title{Schrödinger's Seed: Purr-fect Initialization for an Impurr-fect Universe}

\author{
Mi Chen\inst{1}
\and
Renhao Ye\inst{2,3}
}

\institute{Kapteyn Astronomical Institute, University of Groningen,
Postbus 800, 9700 AV Groningen, The Netherlands
% \email{chenmi@astro.rug.nl}
\and
School of Astronomy and Space Science, University of Chinese Academy of Sciences, 1 East Yanqi Lake Rd., Beijing 100049, P.R. China
\and
Shanghai Astronomical Observatory, Chinese Academy of Sciences, 80 Nandan Rd., Shanghai 200030, China
}

\date{Jokes on April 1, 2026}

% \abstract{}{}{}{}{}
% 5 {} token are mandatory
 
  \abstract
  {Random seed selection in deep learning is often arbitrary — conventionally fixed to values such as 42, a number with no known feline endorsement.}
{We propose that cats, as liminal beings with a historically ambiguous relationship to quantum mechanics, are better suited to this task than random integers.}
{We construct a cat-driven seed generator inspired by the first Friedmann equation, and test it by mapping 21 domestic cats' physical properties — mass, coat pattern, eye colour, and name entropy — via a Monte "Catlo" sampling procedure.}
{Cat-driven seeds achieve a mean accuracy of 92.58\%, outperforming the baseline seed of 42 by $\sim$2.5\%. 
% Fenrir reigns supreme; Paopao does not. 
Cats from astrophysicist households perform marginally better, suggesting cosmic insight may be contagious.}
{The Universe responds better to cats than to arbitrary integers. Whether cats are aware of this remains unknown. 
% catcosmo is available on GitHub and PyPI.
}


   \maketitle

%%%%%%%%%%%%%%%%%%%%%%%%%%%%%%%%%%%%%%%%%%%%%%%%%%%%%%%%%%%%%%
\section{Introduction}
The Universe is, by design, impurr-fect. Quantum fluctuations in the early cosmos seeded the large-scale structure we observe today, giving rise to a magnificent zoo of galaxy morphologies: spirals, ellipticals, lenticulars, mergers, and everything in between \citep{hubble1926extragalactic}. Through this same chain of cosmic imperfection, these structures eventually collapsed into stars, planets, and, after an improbable few billion years, life — including humans, who promptly began classifying the very fluctuations that created them, and cats, who have remained largely indifferent to the endeavour.

With modern surveys from SDSS \citep{york2000sloan} to \textit{Euclid} \citep{laureijs2011euclid} producing data volumes that no human can reasonably process, even with the mood-enhancing assistance of cat-petting. Thus, deep learning has become the galaxy classification tool of choice nowadays \citep{huertas2015catalog, tuccillo2018deep, fluke2020surveying, luo2025galaxy}. Yet in doing so, humanity has introduced a new layer of imperfection: model performance varies with the choice of hyperparameter—particularly the random seed—leading much of the deep learning community to chase state-of-the-art results through unprincipled trial-and-error across seed values.

Cats, by contrast, bring considerably stronger credentials. Although they cannot read galaxy images, throughout history they have been regarded as liminal beings capable of perceiving hidden aspects of reality, and are said to communicate with forces beyond ordinary human perception. While such claims remain outside the scope of empirical validation, \cite{schrodinger_gegenwartige_1935} offered a more rigorous endorsement: a cat in quantum superposition embodies the fundamental uncertainty of reality itself. 

In this paper, we present \texttt{catcosmo}, a random seed generator conditioned on the physical properties of real domestic cats. We argue that cats — as direct products of cosmic fluctuation and time-honoured arbiters of the unknown — are better positioned than 42 to initialise models tasked with classifying the Universe's other impurr-fect creations. For reference, 42 was proposed by \cite{adams1979hitchhiker} as the Answer to the Ultimate Question of Life, the Universe, and Everything, and has since been adopted by the deep learning community without further scrutiny. The cats were not consulted in the writing of this paper. Their seeds were.

%%%%%%%%%%%%%%%%%%%%%%%%%%%%%%%%%%%%%%%%%%%%%%%%%%%%%%%%%%%%%%
\section{Data}
\subsection{Euclid Q1 Data Products}
We make use of imaging data from the \textit{Euclid} Quick Release 1 \citep{collaboration_euclid_2025}. Galaxy stamps of $64\times64$ pixels are extracted from the MER data products, centred on each source. Background-subtracted flux values are clipped at the 99.95th percentile, and an arcsinh stretch with softening parameter $Q=50$ is applied to compress the dynamic range. The resulting images are normalised to the range $[0, 255]$.

\subsection{Galaxy Classification Labels}
Morphological labels are derived from the \textit{Euclid} Galaxy Zoo volunteer classifications \citep{collaboration_euclid_2025-1}, adopting a vote fraction threshold of $>0.6$ to assign discrete labels. We define nine morphological classes: rounded ellipticals, in-between ellipticals, cigar-shaped ellipticals, edge-on disks, face-on spirals, face-on disks without spiral arms, strong-bar galaxies, unbarred galaxies, and mergers. The final labelled sample is divided into training, validation, and test sets in a $6\!:\!2\!:\!2$ ratio using stratified sampling to preserve class balance across splits.

\subsection{cat Parameter Catalogue}
\begin{figure*}

    \centering
    \includegraphics[width=\linewidth]{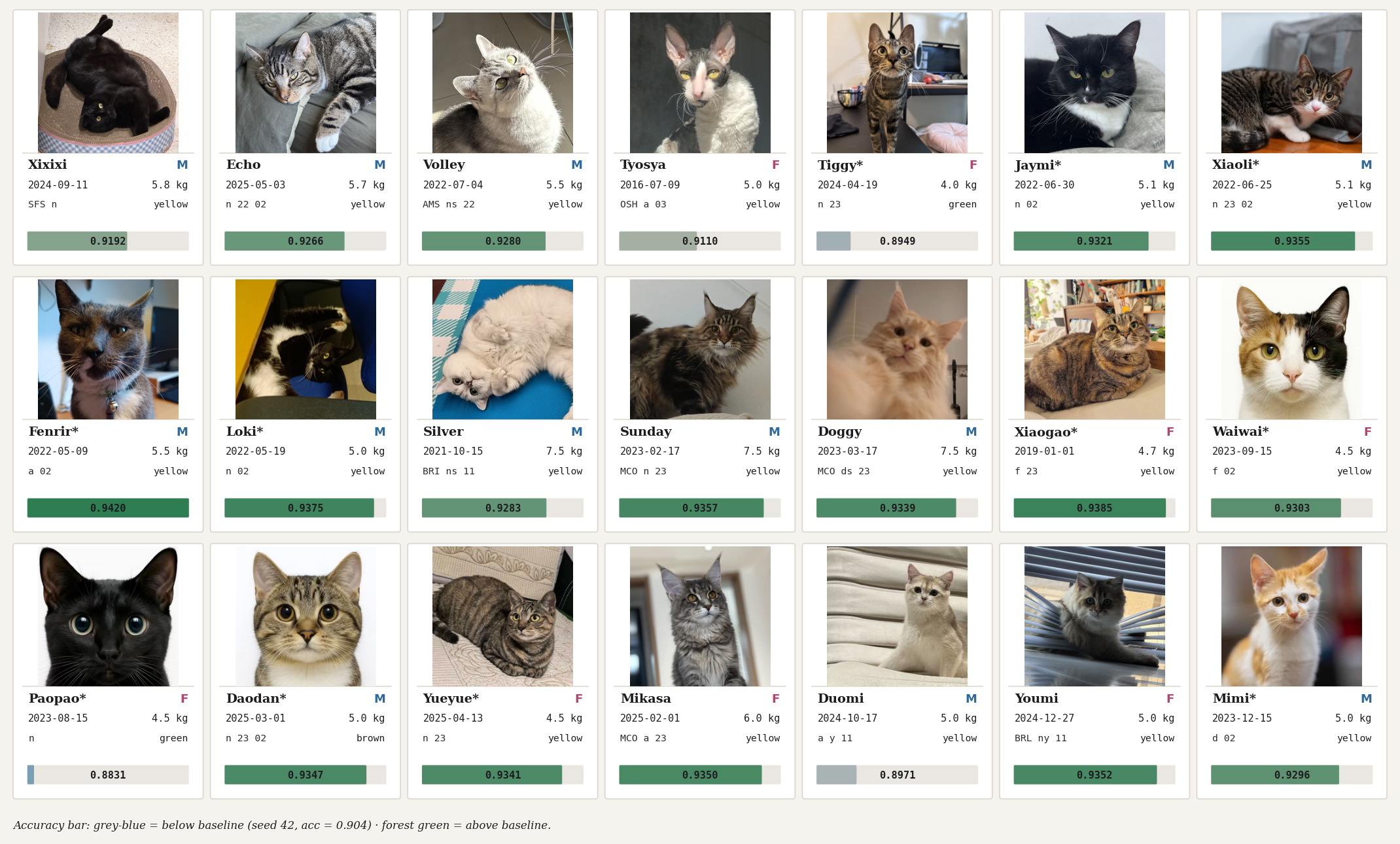}
    \caption{Cat parameter catalogue. Each card displays (from top to bottom): a portrait photograph; the cat’s name (starred entries indicate cats owned by current or former astrophysicists); sex (M/F), date of birth, and body mass; coat pattern code and eye colour; and a colour-coded accuracy bar. The bar colour indicates performance relative to the baseline seed 42 (accuracy = 0.904): darker forest green denotes higher accuracy, while grey-blue denotes lower accuracy. The seed values are generated using the cat-driven Random Seed Generator with Monte "Catlo", based on each cat’s intrinsic parameters. A control entry with the fixed seed 42 is listed separately at the bottom as a baseline reference.}
    
    \label{fig:cat}
    
\end{figure*}

We compiled a sample of 21 domestic cats, for which a set of physical and observational parameters was assembled (See Fig~\ref{fig:cat}). 
The dataset includes a unique Cat ID, date of birth, body weight, biological sex, coat colour pattern (encoded in EMS notation), and eye colour. 
These parameters are chosen to represent both intrinsic properties (e.g. age, mass, and sex) and observable features (e.g. coat pattern and eye colour), enabling a mapping to cosmological quantities within our framework.

\section{Methods}
\subsection{Neural Network}
We adopt a ResNet-50 architecture \citep{he_deep_2015} with two modifications to accommodate our data: The model is trained for 20 epochs from scratch using the Adam optimiser with a learning rate of $10^{-3}$ and a batch size of 256, minimising the cross-entropy loss. The best-performing checkpoint is selected based on validation accuracy.
\subsection{Cat-driven Random Seed Generator with Monte ``Catlo'' }
To help the deep learning (DL) model better ``understand'' the Universe — or at least not completely misunderstand it — we introduce a stochastic random seed generator inspired by the First Friedmann Equation \citep{friedman1922krummung}:
\begin{equation}
\left( \frac{\dot{a}_{\mathrm{cat}}}{a_{\mathrm{cat}}} \right)^2
= H_{\mathrm{cat}}^2
\left[
\Omega_{\mathrm{mass}}\, a_{\mathrm{cat}}^{-3}
+ \Omega_{\mathrm{coat}}\, a_{\mathrm{cat}}^{-2}
+ \Omega_{\Lambda,\mathrm{cat}}
\right].
\end{equation}

Here, $a_{\mathrm{cat}}$ is the cat scale factor describing the expansion state of the cat-conditioned Universe, and $\dot{a}_{\mathrm{cat}}$ its temporal derivative. The parameter $H_{\mathrm{cat}}$ denotes the cat-calibrated Hubble constant, while the density terms $\Omega_{\mathrm{mass}}$, $\Omega_{\mathrm{coat}}$, and $\Omega_{\Lambda,\mathrm{cat}}$ represent contributions from baryonic mass, coat-pattern-induced curvature, and vacuum fluctuations associated with effectively invisible cats, respectively.

The scale factor is assigned according to the sex of the cat, such that $a_{\mathrm{cat}}=1.05$ for male cats and $a_{\mathrm{cat}}=0.95$ for female cats. The Hubble parameter is defined as
\begin{equation}
H_{\mathrm{cat}} = \frac{1}{A_{\mathrm{cat}} + \epsilon},
\end{equation}
where $A_{\mathrm{cat}}$ denotes the age of the cat and $\epsilon=0.1$ is a small regularisation constant introduced to prevent divergences for extremely young cats.

The matter term is defined as the relative mass between the cat and its owner,
\begin{equation}
\Omega_{\mathrm{mass}} = \frac{M_{\mathrm{cat}}}{M_{\mathrm{master}}},
\end{equation}
introducing a scaling with the owner mass. As a result, $\Omega_{\mathrm{mass}}$ is anti-correlated with $M_{\mathrm{master}}$, such that the dynamical importance of the cat increases in systems with lower host mass.

The curvature term is parametrised through the encoded coat colour,
\begin{equation}
\Omega_{\mathrm{coat}} = E_{\mathrm{EMS}},
\end{equation}
where $E_{\mathrm{EMS}}$ denotes the numerical encoding of the cat’s coat pattern based on the EMS (Easy Mind System) classification. The EMS is a standardised coding scheme established by the Fédération Internationale Féline (FIFe)\footnote{\url{https://fifeweb.org/cats/ems-system/}}, widely used to identify cat breeds and coat characteristics.

The dark-cat-energy contribution is defined through the closure relation
\[
\Omega_{\Lambda,\mathrm{cat}} = 1 - \Omega_{\mathrm{mass}} - \Omega_{\mathrm{coat}},
\]
which enforces a normalised energy budget analogous to the standard $\Lambda$CDM framework. Physically, this term represents the fraction of the system not directly observable through mass or coat properties, capturing the effectively invisible component of the cat.

However, the current formulation does not fully capture cat-dependent contributions to the expansion dynamics. To account for this, we introduce a correction term
\begin{equation}
\mathcal{F}_{\mathrm{name}} = 1 + \alpha\,\mathcal{H}_{\mathrm{name}},
\end{equation}
where $\mathcal{H}_{\mathrm{name}}$ is the entropy of the encoded cat name and $\alpha=0.2$ is a tunable parameter.

Finally, to model the intrinsic unpredictability of cats, we incorporate a stochastic component via a Monte ``Catlo'' procedure based on eye colour. Specifically, we introduce a Gaussian perturbation $\epsilon_{\mathrm{eye}} \sim \mathcal{N}(0, \sigma_{\mathrm{eye}})$, where $\sigma_{\mathrm{eye}}$ depends on the eye colour. This term enters as a multiplicative correction, leading to the final seed definition:
\begin{equation}
\label{eq:seed}
\mathrm{Seed}
= \left( \frac{\dot{a}_{\mathrm{cat}}}{a_{\mathrm{cat}}} \right)^2
\, \mathcal{F}_{\mathrm{name}}
\, \left(1 + \epsilon_{\mathrm{eye}}\right).
\end{equation}

\section{Results}
In this section, we train the model using the same architecture and dataset, but with different cat-conditioned random seeds derived from Eq.~\ref{eq:seed}. For comparison, we also train a model using the widely adopted seed value of 42. The resulting accuracies are reported in the colour-coded accuracy bar of Fig~\ref{fig:cat}.

The highest accuracy, corresponding to the most ``clever'' cat in our sample, is achieved by Fenrir, a grey cat partially covered with white patterns. In contrast, the lowest accuracy is obtained for Paopao, a three-year-old black cat, indicating a weaker resonance with the Universe.

Overall, most cat-conditioned seeds outperform the standard seed value of 42. The mean accuracy obtained using cat-conditioned random seeds is $92.58\%$, compared to $90.4\%$ for the baseline seed, suggesting that cat-informed initialisation improves the model's perception of the Universe.

We further compare cats from astrophysicist households with those from non-astrophysicist environments. The average accuracy for cats owned by (former) astrophysicists is $92.63\%$, slightly higher than $92.5\%$ for other cats. This result suggests that the research background of the owner may introduce a mild enhancement in the cat's resonance with the Universe.

\section{Conclusions}
In this study, we present \texttt{catcosmo}, a cat-inspired open-source software package developed to enhance our perception of the universe and improve the performance of galaxy morphology classification tasks. The main findings of this work are summarised as follows:

\begin{enumerate}
  \item Cat-informed random seeds enable a more effective exploration of the optimisation landscape compared to the standard benchmark seed of 42.

    \item We find a slight performance advantage for cats from astrophysical environments, suggesting a possible link between environmental exposure and cosmological sensitivity.
\end{enumerate}

In summary, the Universe appears to favour cats over random integers. Researchers are therefore encouraged to take advantage of their pet cats to support their work. The \texttt{catcosmo} package is freely available on GitHub\footnote{\url{https://github.com/Chenmi0619/catcosmo}}
 and is also distributed via the Python Package Index\footnote{\url{https://pypi.org/project/catcosmo/}}.
%%%%%%%%%%%%%%%%%%%%%%%%%%%%%%%%%%%%%%%%%%%%%%%%%%%%%%%%%%%%%%
\begin{acknowledgements}
We thank Shiyin Shen, Siyuan Chen, Zihao Mu, Anastasia Lialina, Jiaxuan Xi, Ruoxi Zhao, Junjie Liu, Yinting Jin, Floris Prins, Jiayin Lu, Qingyue Yang, Ruqiu Lin and Chunyan Li for providing the cat data and giving advice. 
We thank Yunliang Zheng and Shiyin Shen for advising on the submission time of the April Fools paper. We thank Tiggy for the accompaniment (annoyance) when writing the paper.
This work has made use of the Quick Release (Q1) data from the Euclid mission of the European Space Agency (ESA).
Happy April Fools' Day!
\end{acknowledgements}

\bibliographystyle{aa} 
\bibliography{ref}

% \end{appendix}
\end{document}